\documentclass[3p,times]{elsarticle}

\usepackage{ecrc}

\usepackage{epstopdf}

\volume{00}

\firstpage{1}

\journalname{Nuclear Physics A}

\runauth{Author1 et al.}

\jid{nupha}

\jnltitlelogo{Nuclear Physics A}

\usepackage{graphicx}
\usepackage{amsmath,amssymb}

\begin{document}

\begin{frontmatter}



\title{Measurements of the heavy-flavour nuclear modification factor
in p--Pb collisions at $\sqrt{s_{\rm NN}}$ = 5.02 TeV with ALICE at the LHC}

\author{Shuang Li for the ALICE Collaboration}
\address{Key Laboratory of Quark and Lepton Physics (MOE) and Institute of Particle
Physics,  Central China Normal University, Wuhan, China \\
Laboratoire de Physique Corpusculaire, Clermont Universit\'e,
Universit\'e Blaise Pascal, CNRS-IN2P3, Clermont-Ferrand, France}




\begin{abstract}
The heavy-flavour nuclear modification factor $R_{\rm pPb}$ has been measured with
the ALICE detector in p--Pb collisions at the nucleon--nucleon center of mass
energy $\sqrt{s_{\rm NN}}$ = 5.02 TeV at the CERN LHC in a wide rapidity and transverse
momentum range, as well as in several decay channels.
$R_{\rm pPb}$ is consistent with unity within uncertainties at mid-rapidity and forward rapidity.
In the backward region a slight enhancement of the yield of heavy-flavour decay muons
is found in the region $2<p_{\rm T}<4$ GeV/$c$.
The results are described within uncertainties by theoretical calculations that include
initial-state effects. The measurements add experimental evidence that the
suppression of heavy-flavour production observed at high $p_{\rm T}$ in
central Pb--Pb collisions with respect to pp collisions is
due to a medium effect induced by the interaction of heavy quarks with the partonic matter.
\end{abstract}


\end{frontmatter}


\section{Introduction}
\vspace{-0.7em}
\label{intro}
Heavy quarks (charm and beauty) are essential probes of the properties of the medium created in heavy-ion
collisions, since they are produced in the early stage of hadronic collisions via
scattering processes with large momentum transfer. Heavy-flavour production has been
measured via semi-electronic and semi-muonic decays, as well as fully reconstructed
D mesons, in pp, p--Pb and Pb--Pb collisions with ALICE~\cite{alice} at the LHC.
In central Pb--Pb collisions at $\sqrt{s_{\rm NN}}$ = 2.76 TeV,
a strong suppression of high transverse momentum ($p_{\rm T}$)
D mesons and electrons (muons) from heavy-flavour hadron decays was observed at
mid- (forward) rapidity~\cite{dmesonPbPb, ePbPb, muonPbPb}.
This suppression is interpreted as an effect of parton energy loss in the
medium created in heavy-ion collisions. However, a quantitative understanding
of the Pb-Pb results requires cold nuclear matter (CNM) effects to be taken into account,
which can be accessed by studying p--Pb collisions~\cite{dmesonpPb, jpsipPb} assuming that
a hot and dense extended system is not formed in such collisions. 
CNM effects are studied using the nuclear modification factor $R_{\rm pPb}$, defined as
\begin{equation}
\vspace{-0.4em}
R_{\rm pPb}=\frac{1}{\langle T_{\rm pPb}\rangle}\times\frac{\rm d\it N_{\rm pPb}/ \rm d\it p_{\rm T}}{\rm d\it \sigma_{\rm pp}/\rm d\it p_{\rm T}}
           =\frac{1}{\rm A}\times\frac{\rm d\it \sigma_{\rm pPb}/ \rm d\it p_{\rm T}}{\rm d\it \sigma_{\rm pp}/\rm d\it p_{\rm T}}
\end{equation}
where $\langle T_{\rm AA} \rangle$ is the average nuclear overlap function
estimated through the Glauber model,
which gives $\langle T_{\rm AA} \rangle = {\rm 0.0983\pm0.0035} \rm mb^{-1}$~\cite{identpPb};
$\rm d\it N_{\rm pPb}/\rm d\it p_{\rm T}$ ($\rm d\it \sigma_{\rm pPb}/\rm d\it p_{\rm T}$) is
the $p_{\rm T}$-differential yield (cross section) in p--Pb collisions and
$\rm d\it \sigma_{\rm pp}/\rm d\it p_{\rm T}$ is
the $p_{\rm T}$-differential cross section in pp collisions;
A is the mass number of the Pb nucleus.
The value of $R_{\rm pPb}$ is unity in absence of nuclear effects.

\vspace{-0.5em}
\section{Heavy-flavour measurements with ALICE}
\vspace{-0.7em}
ALICE is the dedicated heavy-ion experiment at the LHC.
It allows to investigate heavy-flavour production in several decay channels over a
wide rapidity and transverse momentum range.
The ALICE detector consists of a set of central barrel detectors ($\mid\eta_{\rm lab}\mid< 0.9$),
a muon spectrometer ($-4.0<\eta_{\rm lab}<-2.5$) and global detectors
for triggering and event characterization purposes.
At mid-rapidity, the Inner Tracking System (ITS) and
the Time Projection Chamber (TPC) provide track reconstruction 
down to very low transverse momentum ($\sim$100 MeV/$c$)
with a momentum resolution better than 4$\%$ for $p_{\rm T}<20$ GeV/$c$,
as well as good impact parameter
(distance of closest approach of the track to the primary interaction vertex)
resolution~\cite{alice, cbperform}. 
D mesons are reconstructed via their hadronic decay channels
and electrons from semileptonic decays of charm and$/$or beauty hadrons are measured.
At forward rapidity, muons from
heavy-flavour hadron decays can be measured, since the ALICE muon spectrometer allows to identify muons
by requiring that a track reconstructed in the tracking system is matched with
the corresponding track candidate in the muon trigger system.

D mesons are reconstructed
from their hadronic decay channels 
${\rm D^{0}}$ $\rightarrow$ ${\rm K^{-}\pi^{+}}$ (BR = 3.88$\%$),
${\rm D^{+}}$ $\rightarrow$ ${\rm K^{-}\pi^{+}\pi^{+}}$ (BR = 9.13$\%$),
${\rm D^{*+}}$ $\rightarrow$ ${\rm D^{0}\pi^{+}}$ $\rightarrow$ ${\rm K^{-}\pi^{+}\pi^{+}}$ (BR = 67.7$\%$)
and ${\rm D^{+}_{s}}$ $\rightarrow$ ${\rm \phi\pi^{+}}$ $\rightarrow$ ${\rm K^{+}K^{-}\pi^{+}}$ (BR = 2.28$\%$).
The selection of D meson
candidates against the large combinatorial background is based on the
reconstruction of decay vertices displaced by a few hundred $\mu m$ from the
interaction vertex, exploiting the $c\rm \tau$ of 
${\rm D^{0}}$, ${\rm D^{+}}$ and ${\rm D^{+}_{s}}$, which is
about $\rm 123-300~\mu m$ depending on the D-meson species.
In order to further enhance the ratio between the D-meson
signal and the combinatorial background,
the measurements of the particle time-of-flight from the collision point to the Time Of Flight (TOF) detector
and of the specific energy loss in the TPC gas are used to identify
$\rm K^{\pm}$ and $\pi^{\pm}$~\cite{dmesonhp}.
The measurement is performed in the rapidity interval $-0.96<y_{\rm cms}<0.04$
(which is specific for p-Pb collisions) over a wide transverse momentum range.

The electrons are identified with TPC and TOF at low $p_{\rm T}$,
as well as the TPC and Electromagnetic Calorimeter (EMCAL) at high $p_{\rm T}$.
The measurement of electrons from heavy-flavour hadron decays at mid-rapidity
($-1.06<y_{\rm cms}<0.14$)
requires the subtraction of contributions from several background sources from the inclusive
electron distribution. The dominant contribution is from photon conversions
in the detector material and Dalitz decays of light neutral mesons
($\pi^{\rm 0}$ and $\eta$, mainly).
These contributions are statistically subtracted by using
the cocktail
(i.e. a calculation of the background based mainly on measured
$p_{\rm T}$-differential cross sections of the main electron sources)
and invariant mass
(i.e. the measurement of electrons from photon conversions and Dalitz
decays via low-mass electron-positron pairs) methods~\cite{ePbPb}.
With the high spatial resolution of the track impact parameter measurement,
one can isolate the beauty-decay contribution
thanks to the much larger lifetime of beauty hadrons ($c\rm \tau \approx {\rm 500\mu m}$)
compared to that of charm hadrons and other background sources. The electrons from their
semileptonic decays have, consequently, a larger average impact parameter with respect
to the primary interaction vertex.

Muons from heavy-flavour hadron decays have been measured at forward
(i.e. proton-beam direction, $2.5<y_{\rm cms}<3.53$) and backward
(i.e. Pb-beam direction, $-4<y_{\rm cms}<-2.96$) rapidity, respectively,
by analyzing data collected with different beam configurations.
The main source of background in the $p_{\rm T}$-differential
inclusive spectrum consists of muons from light hadron decays
($\pi^{\pm}$ and $\rm K^{\pm}$, mainly).
This contribution dominates the muon yield for
$p_{\rm T}<2$ GeV/$c$ and prevents a measurement of muons from heavy-flavour
decays at low $p_{\rm T}$.
At larger $p_{\rm T}$ the contamination is estimated via
Monte-Carlo simulations, as well as with a data-driven method based
on the extrapolation of charged hadron
yields~\cite{cbhadron} measured at mid-rapidity with ALICE to forward rapidity.

\vspace{-0.5em}
\section{Results}
\begin{figure}[htbp]
\vspace{-1.1em}
\begin{center}
\includegraphics*[width=0.38\textwidth]{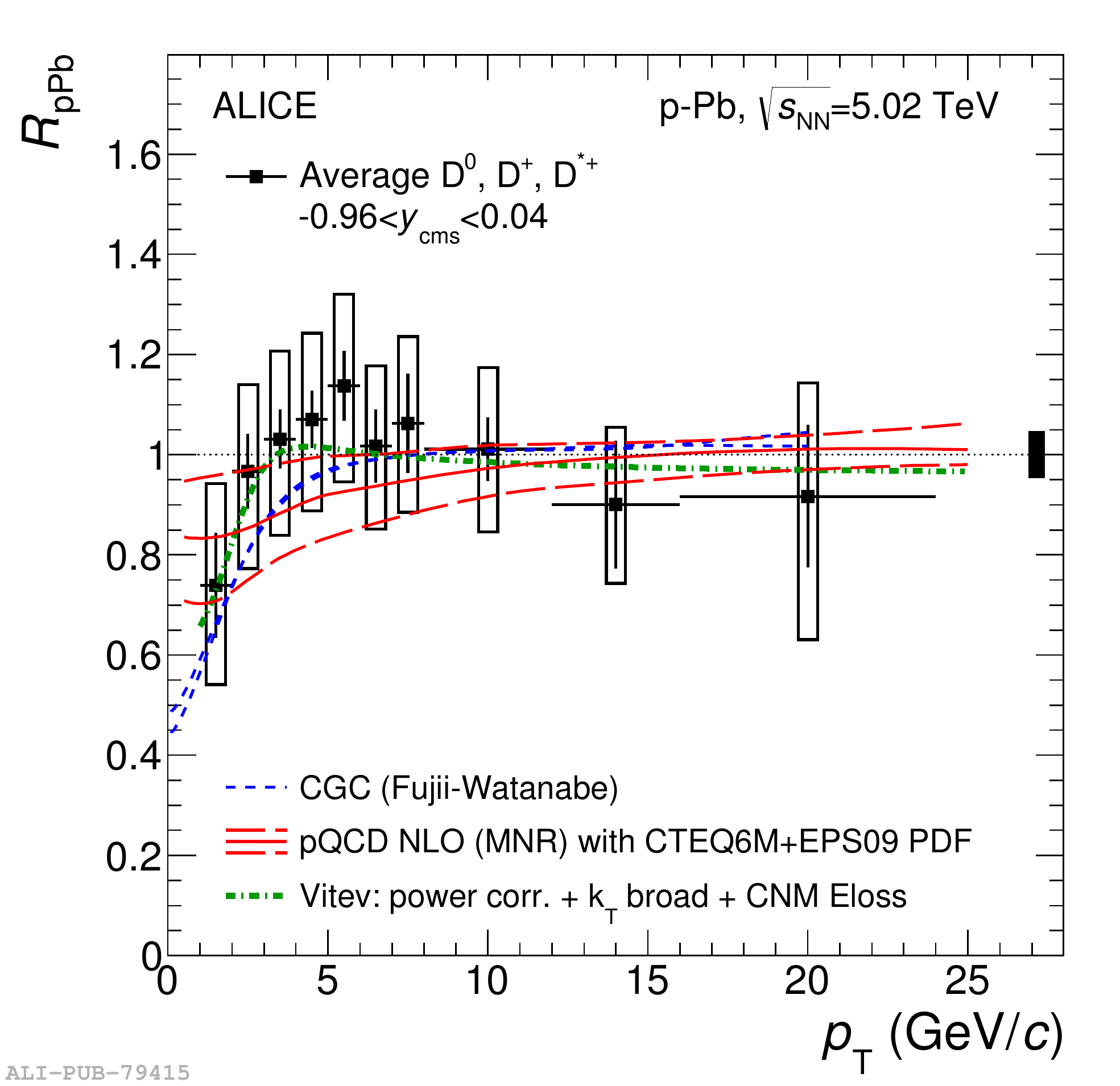}
\hspace{1.0em}
\includegraphics*[width=0.38\textwidth]{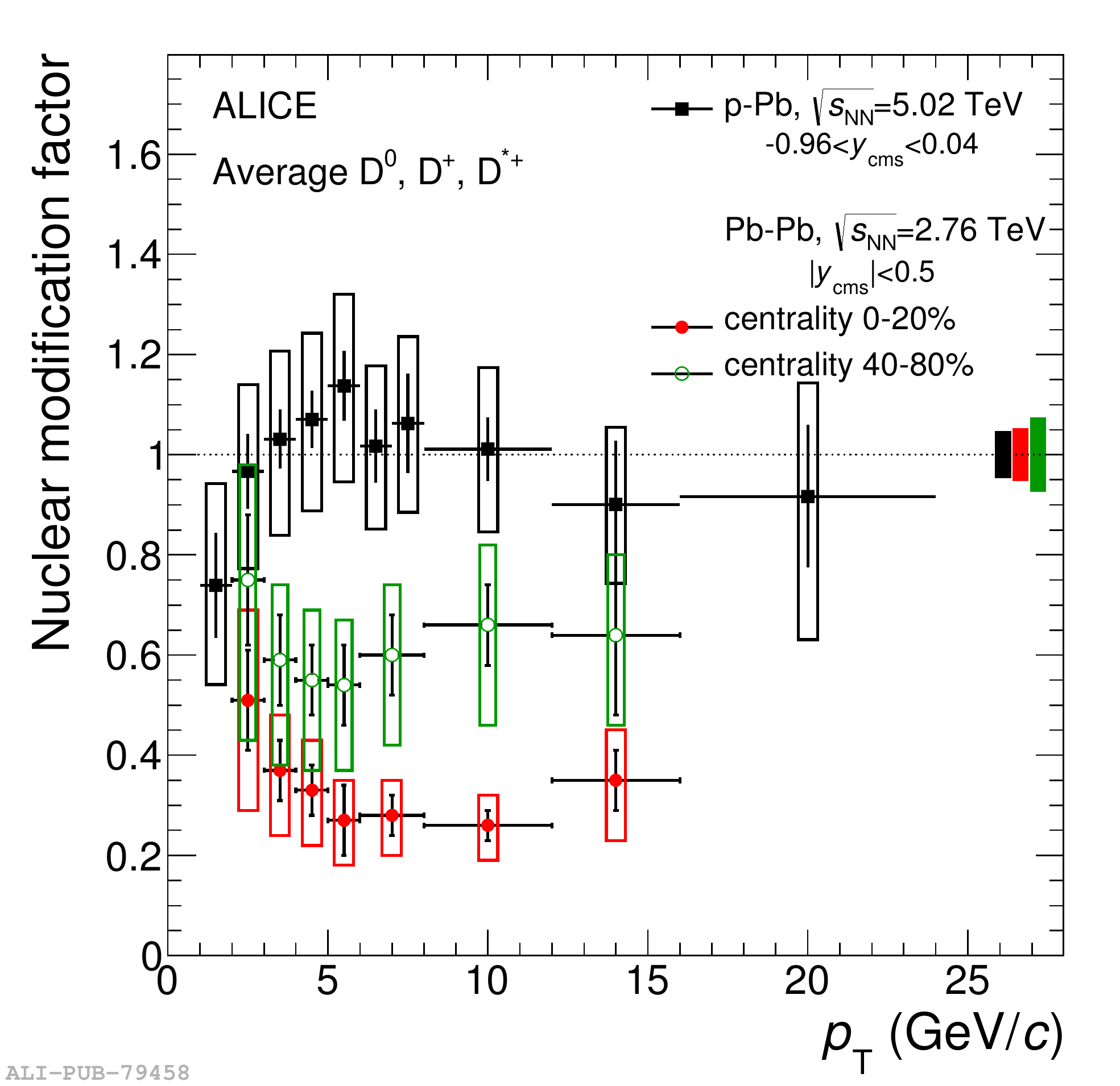}
\end{center}
\vspace{-1.7em}
\caption{Left: Average $R_{\rm pPb}$ of prompt ${\rm D^{0}}$, ${\rm D^{+}}$ and
        ${\rm D^{*+}}$ mesons as a function of $p_{\rm T}$ compared to model
	calculations~\cite{MNR, EPS09, CTEQ6M, CGC, kTBroad};
        Right: Average $R_{\rm pPb}$ of prompt ${\rm D^{0}}$, ${\rm D^{+}}$ and
        ${\rm D^{*+}}$ mesons as a function of $p_{\rm T}$ compared to the prompt D-meson $R_{\rm AA}$ in central
	(0-20$\%$) and semi-peripheral (40-80$\%$) Pb--Pb collisions at $\sqrt{s_{\rm NN}}$ = 2.76 TeV~\cite{ePbPb}.}
\label{fig:DMeson}
\end{figure}

\vspace{-0.7em}
The nuclear modification factors, $R_{\rm pPb}$, of ${\rm D^{0}}$, ${\rm D^{+}}$, ${\rm D^{*+}}$
and ${\rm D^{+}_{s}}$ mesons are consistent with each other, and they are compatible with unity
within uncertainties~\cite{dmesonpPb}.
The pp reference cross section at $\sqrt{s}$ = 5.02 TeV is
obtained by a pQCD-based energy scaling~\cite{enscal,FONLL} of the $p_{\rm T}$-differential cross
sections measured at $\sqrt{s}$ = 7 TeV~\cite{dmesonpp}.
Figure~\ref{fig:DMeson} (left) presents the average $R_{\rm pPb}$
of prompt ${\rm D^{0}}$, ${\rm D^{+}}$ and ${\rm D^{*+}}$ in $1<p_{\rm T}<24$ GeV/$c$.
The $R_{\rm pPb}$ can
be described by means of next-to-leading-order (NLO) pQCD~\cite{MNR} calculations, including the
EPS09~\cite{EPS09} nuclear modification of the CTEQ6M~\cite{CTEQ6M} Parton Distribution Functions (PDF)
and calculations based on the Color Glass Condensate (CGC~\cite{CGC}) initial-state prescription.
The data are also well described by calculations which include
energy loss in cold nuclear matter,
nuclear shadowing and $k_{\rm T}$-broadening~\cite{kTBroad}.
As demonstrated in Figure~\ref{fig:DMeson} (right),
cold nuclear matter effects are small for $p_{\rm T}\gtrsim$3 GeV/$c$.
This indicates that the suppression of D mesons observed in the 20$\%$ most
central Pb--Pb collisions at $\sqrt{s_{\rm NN}}$ = 2.76 TeV~\cite{dmesonPbPb} results from
an effect due to the presence of the hot and dense medium.

\begin{figure}[htbp]
\vspace{-1.7em}
\begin{center}
\includegraphics*[width=0.40\textwidth]{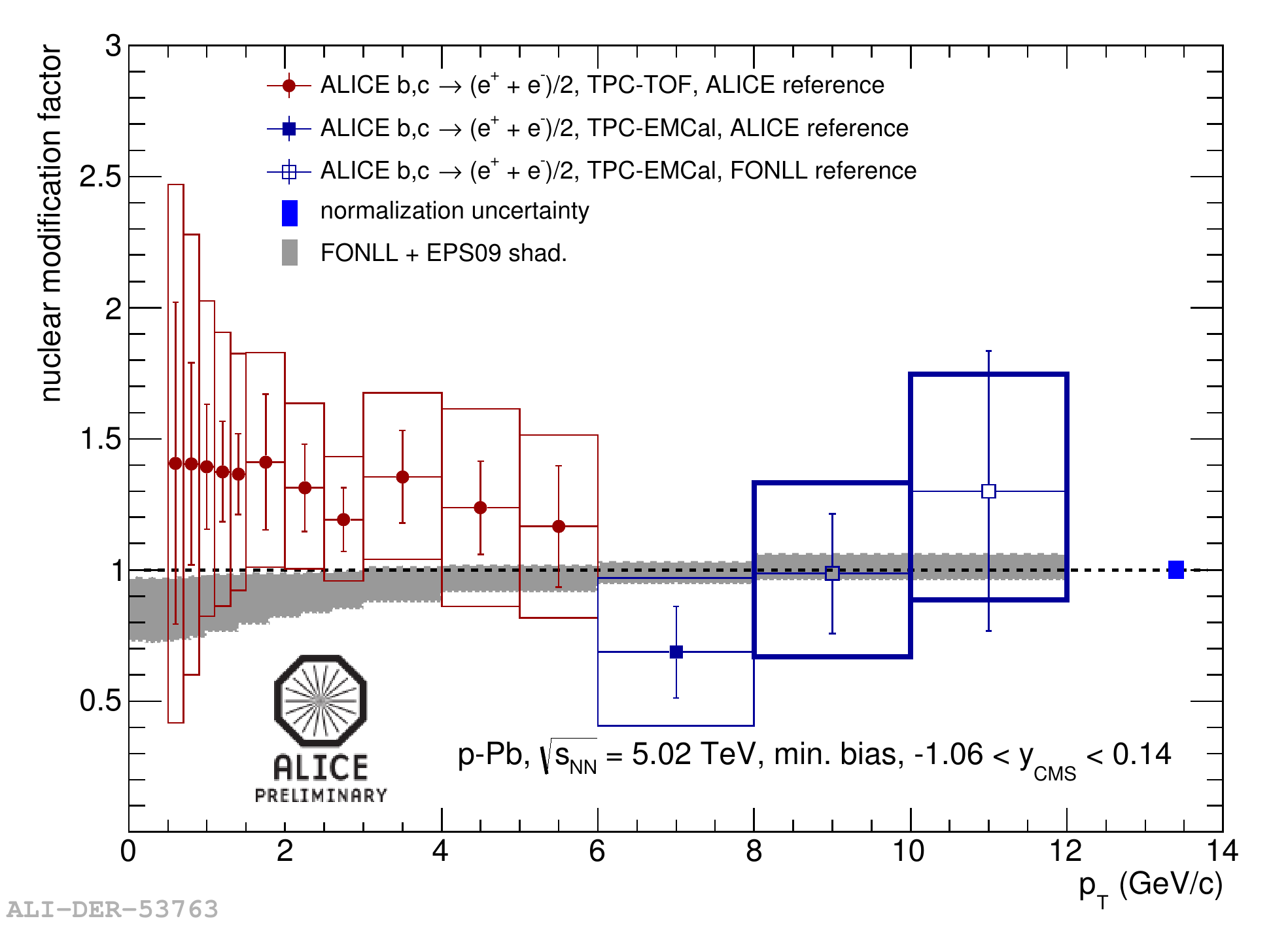}
\includegraphics*[width=0.40\textwidth]{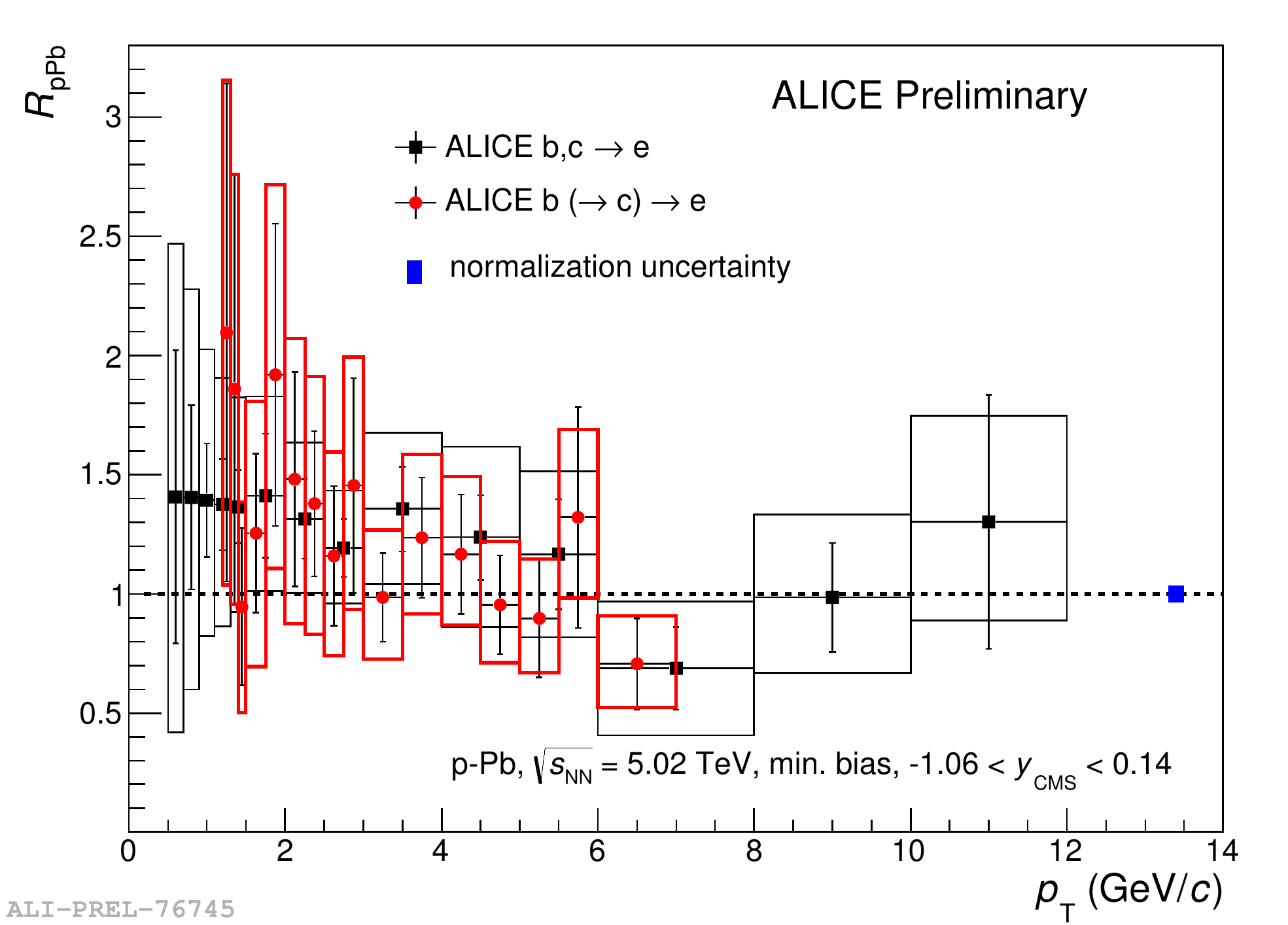}
\end{center}
\vspace{-1.7em}
\caption{Left: $R_{\rm pPb}$ of electrons from heavy-flavour hadron decays
        as a function of $p_{\rm T}$ compared to model calculations~\cite{FONLL, EPS09};
        Right: $R_{\rm pPb}$ of electrons from heavy-flavour hadron decays
        compared to $R_{\rm pPb}$ of electrons from beauty-hadron decays.}
\label{fig:Electron}
\end{figure}

The $R_{\rm pPb}$ of electrons from heavy-flavour hadron decays in p--Pb collisions
at $\sqrt{s_{\rm NN}}$ = 5.02 TeV is shown in Figure~\ref{fig:Electron} (left).
The pp reference is obtained following a strategy similar to the one used for the D-meson analysis. 
The $R_{\rm pPb}$ is compatible with unity, showing that cold nuclear matter effects are small. 
The measurement is compared with a pQCD calculation with the EPS09~\cite{EPS09} NLO
parameterization of nuclear PDFs. Within uncertainties,
the data can be described by the model predictions.
Figure~\ref{fig:Electron} (right) displays the measured $R_{\rm pPb}$ of electrons from
heavy-flavour hadron decays together with that of electrons from beauty-hadron decays.
The $R_{\rm pPb}$ of electrons from beauty-hadron decays is consistent with unity within
uncertainties, and it is also compatible with that of electrons from heavy-flavour hadron decays.
The data indicate that the suppression of electrons from heavy-flavour hadron decays observed in the 10$\%$ most
central Pb--Pb collisions at $\sqrt{s_{\rm NN}}$ = 2.76 TeV~\cite{ePbPb} results from
a hot medium effect.

\begin{figure}[htbp]
\vspace{-1.1em}
\begin{center}
\includegraphics*[width=0.45\textwidth]{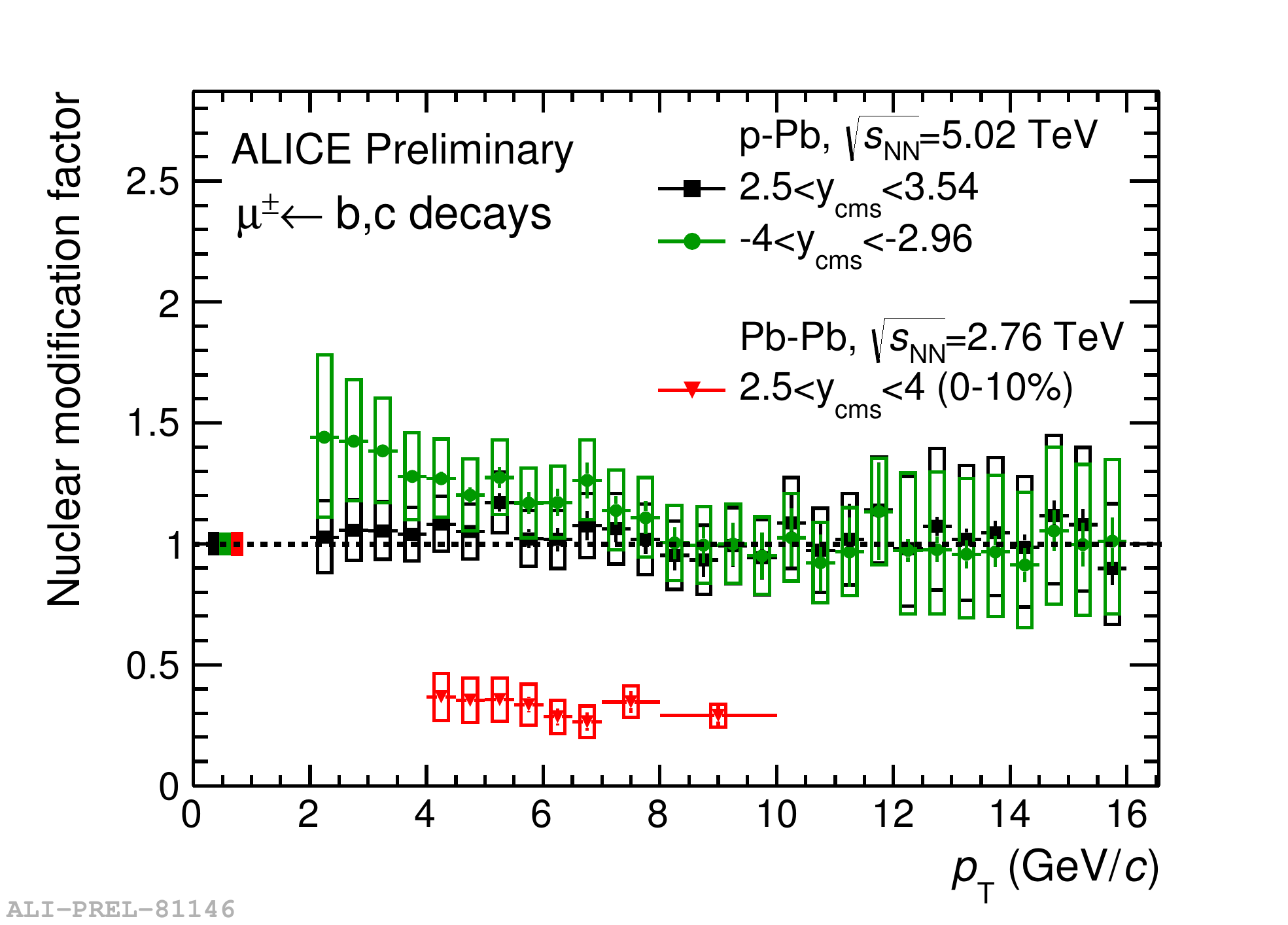}
\includegraphics*[width=0.45\textwidth]{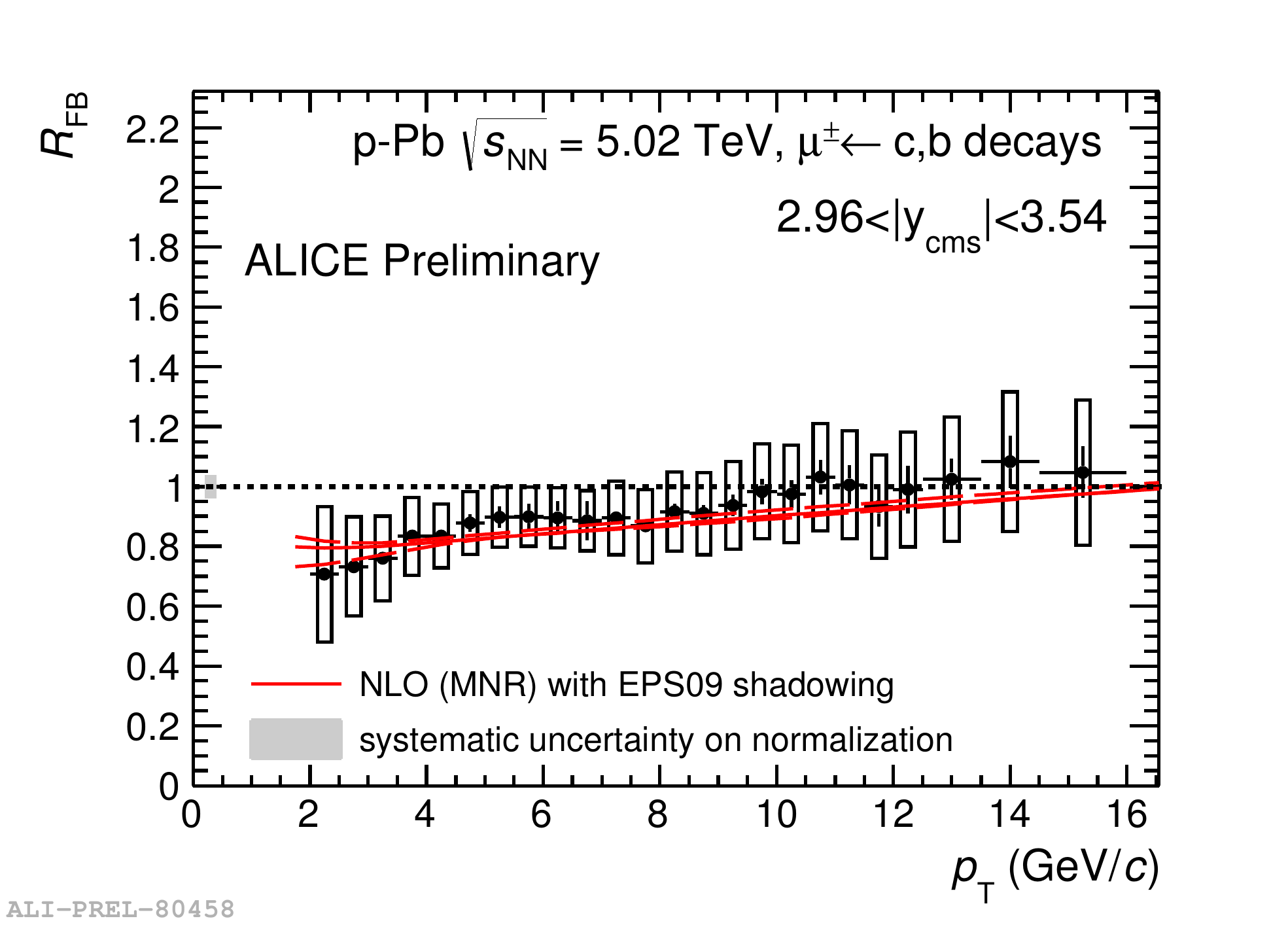}
\end{center}
\vspace{-1.7em}
\caption{Left: $R_{\rm pPb}$ of muons from heavy-flavour hadron decays
        as a function of $p_{\rm T}$ at forward rapidity ($2.5<y_{\rm cms}<3.53$) and
        backward rapidity ($-4<y_{\rm cms}<-2.96$), compared to the $R_{\rm AA}$ for
        muons from heavy-flavour hadron decays in the 10$\%$ most central Pb--Pb
        collisions at $\sqrt{s_{\rm NN}}$ = 2.76 TeV~\cite{muonPbPb};
        Right: Forward-to-backward ratio $R_{\rm FB}$ of muons from heavy-flavour hadron decays
        compared to a model calculation~\cite{MNR, EPS09}.}
\label{fig:Muon}
\end{figure}

The nuclear modification factor of muons from heavy-flavour hadron
decays, shown in Figure~\ref{fig:Muon} (left), is measured in the center
of mass rapidity domains $2.5<y_{\rm cms}<3.53$ (forward rapidity) and
$-4<y_{\rm cms}<-2.96$ (backward rapidity).
The pp reference cross section of heavy-flavour hadron decay muons at $\sqrt{s}$ = 5.02 TeV is
obtained via a pQCD-based energy scaling procedure~\cite{FONLL}
(i.e. calculated taking as input the $p_{\rm T}$-differential cross
sections measured at $\sqrt{s}$ = 7 TeV~\cite{muonpp}), performed at forward and backward
rapidity, respectively. 
This procedure is used to obtain the $p_{\rm T}$-differential cross section up to 12 GeV/$c$.
In order to measure the $R_{\rm pPb}$ in a wider transverse
momentum range, the $p_{\rm T}$-differential cross section in pp collisions at $\sqrt{s}$ = 5.02 TeV was
extrapolated to higher $p_{\rm T}$
using the spectrum predicted by FONLL scaled to match the obtained pp reference in the range $6<p_{\rm T}<12$ GeV/$c$.
The $R_{\rm pPb}$ measured at forward rapidity is consistent with
unity within uncertainties. $R_{\rm pPb}$ at backward rapidity
is slightly larger than unity in the range $2<p_{\rm T}<4$ GeV/$c$
and close to unity at higher $p_{\rm T}$.
This supports the previous observation that the suppression measured in the 10$\%$ most central Pb--Pb
collisions at $\sqrt{s_{\rm NN}}$ = 2.76 TeV~\cite{muonPbPb} results from hot and dense medium effect.  

Another way to quantify cold nuclear matter effects is to study the forward-to-backward ratio
defined as the cross section measured at forward rapidity with respect to the one at backward rapidity,
The main advantage of using such a ratio is that the pp reference and the nuclear overlap
function cancel. The drawback of this approach is the limited statistics,
because the common $y_{\rm cms}$ interval covered at both forward and backward rapidity
is small ($\sim 0.57$ units).
\begin{equation}
R_{\rm FB}(2.96<\mid y_{\rm cms}\mid<3.54)=\frac{\rm d\it \sigma^{\rm forward}/  \rm d\it p_{\rm T}{(\rm 2.96}<y_{\rm cms}{\rm <3.54)}}
                                                {\rm d\it \sigma^{\rm backward}/ \rm d\it p_{\rm T}{(\rm -3.54}<y_{\rm cms}{\rm <-2.96)}}
\end{equation}
The measured $R_{\rm FB}$ is presented in Figure~\ref{fig:Muon} (right).
It is systematically smaller than unity in the range $2<p_{\rm T}<4$ GeV/$c$ and close to unity at higher $p_{\rm T}$.
This is well reproduced by a pQCD calculation including the EPS09~\cite{EPS09} NLO
parameterization of nuclear PDFs~\cite{MNR} in the range $2<p_{\rm T}<16$ GeV/$c$.
The $R_{\rm pPb}$ measured at backward rapidity is slightly underestimated by
such theoretical calculations at low $p_{\rm T}$.

\vspace{-1.0em}
\section{Conclusions}
\vspace{-0.7em}
The nuclear modification factor of heavy-flavour production has been measured
in a wide rapidity and transverse momentum range, as well as in several
decay channels, in p--Pb collisions at
$\sqrt{s_{\rm NN}}$ = 5.02 TeV with ALICE. The $R_{\rm pPb}$ results provide evidence
that cold nuclear matter effects are small, showing that the strong suppression observed in central
Pb--Pb collisions at $\sqrt{s_{\rm NN}}$ = 2.76 TeV is a final state effect due to in-medium parton energy loss.

\vspace{-0.5em}
\section*{Acknowledgments}
\vspace{-0.7em}
{\small
This work is supported partly by the Chinese Ministry of Science and Technology
973 grant 2013CB837803, the NSFC Grant 11375071 and IRG11221504,
CCNU Key grant CCNU13F026, and Key Laboratory QLPL2014P0109.
}






\vspace{-0.5em}
\begin{thebibliography}{00}
\vspace{-0.5em}

\bibitem{alice} K.~Aamodt~{\it et~al.} [ALICE Collaboration], JINST~{\bf 3}, S08002 (2008)
\bibitem{dmesonPbPb} B.~Abelev~{\it et~al.} [ALICE Collaboration], JHEP~{\bf 09}, 112 (2012)
\bibitem{ePbPb} E.~Pereira de Oliveira Filho for the ALICE Collaboration, arXiv:1404.3983 
\bibitem{muonPbPb} B.~Abelev~{\it et~al.} [ALICE Collaboration], Phys.~Rev.~Lett.~{\bf 109}, 112301 (2012)
\bibitem{dmesonpPb} B.~Abelev~{\it et~al.} [ALICE Collaboration], arXiv:1405.3452
\bibitem{jpsipPb} B.~Abelev~{\it et~al.} [ALICE Collaboration], JHEP~{\bf 02}, 073 (2014)
\bibitem{identpPb} B.~Abelev~{\it et~al.} [ALICE Collaboration], Phys. Rev. Lett.~{\bf 110}, 082302 (2013)
\bibitem{cbperform} B.~Abelev~{\it et~al.} [ALICE Collaboration], J.Phys. G~{\bf 30}, 1517 (2004),
                    J.Phys. G~{\bf 32}, 1295 (2006)
\bibitem{dmesonhp} A.~Rossi for the ALICE Collaboration, to be published in Nuclear Physics:
                    proceedings of Hard Probe conference
\bibitem{cbhadron} A.~O.~Velasquez for the ALICE Collaboration, to be published in Nuclear Physics:
		proceedings of Hard Probe conference 
\bibitem{enscal} R.~Averbeck, N.~Bastid, Z.~Conesa del Valle, P.~Crochet, A.~Dainese, X.~Zhang, arXiv:1107.3243
\bibitem{dmesonpp} B.~Abelev~{\it et~al.} [ALICE Collaboration], JHEP~{\bf 01}, 112 (2012)
\bibitem{FONLL} M.~Cacciari~{\it et al.}, JHEP~{\bf 10}, 137 (2012)  
\bibitem{MNR} M.~L.~Mangano~{\it et~al.}, Nucl.~Phys. B~{\bf 373}, 295 (1992)
\bibitem{EPS09} K.~Eskola~{\it et~al.} JHEP~{\bf 0904}, 065 (2012)
\bibitem{CTEQ6M} D.~Stump, J.~Huston, J.~Pumplin, W.~K.~Tung, H.~L.~Lai, S.~Kuhlmann and J.~F.~Owens, JHEP~{\bf 03}, 046 (2003)
\bibitem{CGC} H.~Fujii and K.~Watanabe, Nucl. Phys. A~{\bf 920}, 78 (2013) and Nucl. Phys. A~{\bf 915}, 1 (2013)
\bibitem{kTBroad} R.~Sharma~{\it et~al.}, Phys.~Rev. C~{\bf 80}, 054902 (2009)
\bibitem{muonpp} B.~Abelev~{\it et~al.} [ALICE Collaboration], Phys.~Lett. B~{\bf 708}, 265 (2012)
\end{thebibliography}



\end{document}